\documentclass[5p]{elsarticle}
\usepackage{graphics}
\usepackage{amsmath, amssymb}

\usepackage[colorlinks,linkcolor=blue,citecolor=blue]{hyperref}

\newcommand{\second}{\,\text{s}}
\newcommand{\nM}{\,\text{nM}}

\newcommand{\nMs}{\nM\cdot\second}

\begin{document}
%%%%%%%%%%%%%%%%%%%%%%
% frontmatter                                                                       %
%%%%%%%%%%%%%%%%%%%%%%
\begin{frontmatter}

\title{Gene circuit designs for noisy excitable dynamics}
\author[upc]{Pau Ru\'{e}}
\author[upc]{Jordi Garcia-Ojalvo\corref{cor}}
\ead{jordi.g.ojalvo@upc.edu}
\cortext[cor]{Corresponding author}
\address[upc]{Departament de F\'isica i Enginyeria Nuclear,
Universitat Polit\`ecnica de Catalunya, Edifici GAIA, Rambla de Sant Nebridi s/n, Terrassa 08222,Barcelona, Spain}

\begin{abstract}
Certain cellular processes take the form of activity pulses that can be interpreted in terms of noise-driven excitable
dynamics. Here we present an overview of different gene circuit architectures that exhibit excitable pulses of protein
expression, when subject to molecular noise. Different types of excitable dynamics can occur depending on the
bifurcation structure leading to the specific excitable phase-space topology. The bifurcation structure is not, however,
linked to a particular circuit architecture. Thus a given gene circuit design
can sustain different classes of excitable dynamics depending on the system parameters.
\end{abstract}

\end{frontmatter}

%%%%%%%%%%%%%%%%%%%%%%%%%%%%%%%%%%%%%%%%%%%%%%%%%%%%%%%%%%%%%%%%%%%%%%%%%%%%%%%%%%%%%
\section{Introduction}\label{sec:intro} 

\begin{figure}[htb]
\centering
\centerline{\includegraphics[width=0.3\textwidth]{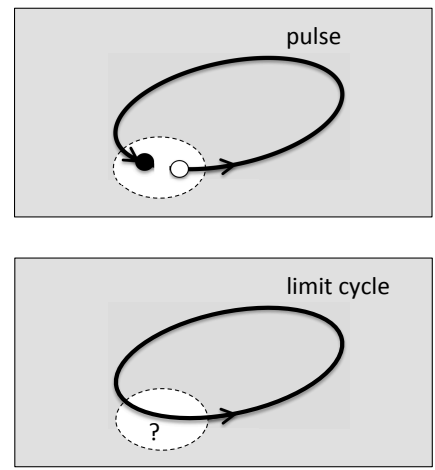}}
\caption{\label{fig:scheme}
Schematic representation of excitable dynamics (top) and its associated limit-cycle behavior (bottom). 
The solid circle in the top plot represents the stable fixed point of the system. 
In this regime, large enough perturbations (starting at the empty circle in top panel) result in a large excursion through phase space before coming back to the stable state.
Representation adapted from \cite{Izhikevich2006}.}
\end{figure}

Dynamical behavior is ubiquitous in gene regulatory processes. While many of the dynamical phenomena exhibited
by genetic circuits take the form of periodic oscillations, in certain cases the behavior is governed by randomly occurring
pulses of protein expression. Mathematically, this type of dynamics can be understood as an instance of {\em excitability},
by which a dynamical system with a stable fixed point is forced to undergo, when subject to a relatively small
perturbation, a large excursion in phase space before relaxing back to the fixed point \cite{revex2004}
(top panel in Fig.~\ref{fig:scheme}).
This feature arises when the system is close to a bifurcation point beyond which the dynamics has the form of a limit cycle
(bottom panel in Fig.~\ref{fig:scheme}).

\begin{figure}[htb]
\centering
\centerline{\includegraphics[width=0.4\textwidth]{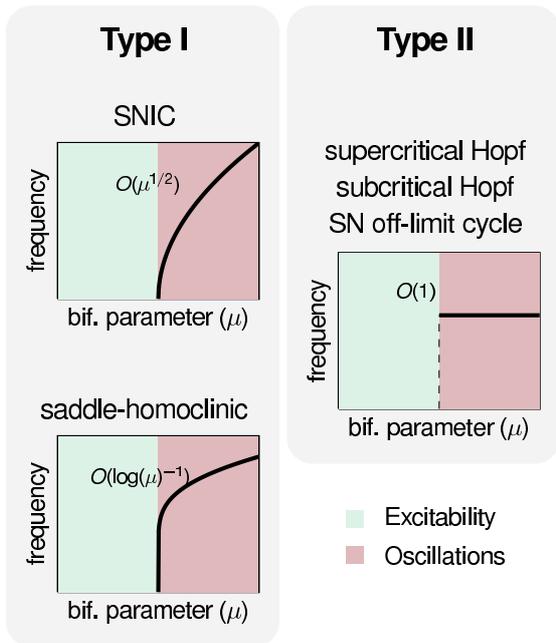}}
\caption{\label{fig:FIrelation}
Frequency response to the increase of a given control parameter, for the two types of excitable behavior
discussed in the text,
as a bifurcation from excitability to oscillations is crossed. The different bifurcation types underlying
the different scenarios are listed in top of each plot.}
\end{figure}

Functionally, excitability provides cells with a mechanism to amplify molecular noise and transform it into a
macroscopic cellular response. Biochemical activity pulses have been reported to exist in
cAMP signaling in amoebae \cite{Hofer:1995uq}, response to DNA damage in human cells \cite{Lahav:2004fk},
competence in bacteria \cite{Suel2006,Suel2007},
and differentiation priming in embryonic stem cells \cite{Kalmar:2009kx}. In a different (but still
biological) context, electrical activity pulses
in a specific cell type, namely neurons, have long been associated with excitable dynamics
\cite{Izhikevich2006}. Within that context, early work by Hodgkin \cite{Hodgkin:1948vn} identified
two main different types of excitability on the basis of the frequency response of the neuron across
a bifurcation leading from excitable to oscillatory behavior (Fig.~\ref{fig:FIrelation}). {\em Type I excitability}
corresponds to a situation in which the limit cycle is born at the bifurcation with a frequency equal to zero,
i.e. the oscillatory activity pulses become infinitely sparse as the bifurcation is approached 
(left panels in Fig.~\ref{fig:FIrelation}). In {\em type II excitability}, on the other hand, the limit cycle is
born with a non-zero frequency (and thus with a finite period), as shown in the right panel of
Fig.~\ref{fig:FIrelation}. 
The distinction is not purely academic: given that neurons encode
information mainly in the timing between pulses, the two types of excitability correspond to two fundamentally
different modes of information transmission \cite{St-Hilaire:2004ys}. Additionally, they exhibit distinct statistical features in their
response to noise \cite{St-Hilaire:2004ys}.

The two types of excitability shown in Fig.~\ref{fig:FIrelation} are associated with different classes of bifurcation
to oscillatory behavior. Two kinds of bifurcation lead to type I excitability.
The first is a saddle-node on an invariant circle (SNIC) bifurcation, in which a stable node and a saddle point appear
together on a limit cycle. In that regime, the trajectory along the remnant of the limit cycle delimits the
excitable pulse. The second bifurcation leading to type I excitability is a saddle-homoclinic bifurcation,
which occurs when a stable limit cycle collides with a saddle point, being transformed into a homoclinic
orbit at the bifurcation point. Again this orbit delimits the excitable pulse. In both
cases the existence of a saddle provides the system with a well defined excitability threshold,
given by the stable manifold of the saddle. The two bifurcation scenarios
are characterized by qualitatively distinct scaling relations of the frequency with respect to the control
parameter \cite{Strogatz1994}, as specified inside the plots at the left column of Fig.~\ref{fig:FIrelation}.

Type II excitability,
on the other hand, is associated with a Hopf bifurcation, either supercritical or subcritical,
or to a saddle-node bifurcation occurring near (not on top of) a limit cycle. In the latter
case, the saddle separates the stable fixed point from the periodic orbit, and again its stable
manifold constitutes a separatrix that determines the excitability threshold. In the case of the
Hopf bifurcation, on the other hand, it is not possible to define in a quantitative way an excitation
threshold in phase space.

Given that the type of excitability determines the statistics of the noise-driven pulses, it is important to
establish, in the context of gene regulation dynamics, the relationship between the gene circuit
architecture and the phase-space topology resulting in excitability. A previous study
\cite{Guantes:2006p1216} has linked
the specific dynamical features of two-component genetic oscillators to the particular molecular
implementation of one of the regulatory links, associating transcriptional regulation with
type I dynamics, and post-translational regulation with type II. Here we show that both types
of excitability are possible in both architectures, and add a third one that has been identified
in bacteria, for which again both types of excitability are possible. Thus, our results indicate that
circuit architecture does in principle not constrain the type of dynamics that the system exhibits.
The three circuits under consideration are described in the next Section, followed in the subsequent
Sections by detailed
bifurcation analyses of the three systems for two parameter sets each, which lead to the two types
of dynamics described above.  
Finally, in Section~\ref{sec:paramspace} we explore the parameter space to confirm that both 
types of excitable behaviors can be achieved for a wide range of parameter values.

\section{A catalogue of excitable circuits}\label{sec:models}

\begin{figure*}[hbt]
\centering
\centerline{\includegraphics[width=0.95\textwidth]{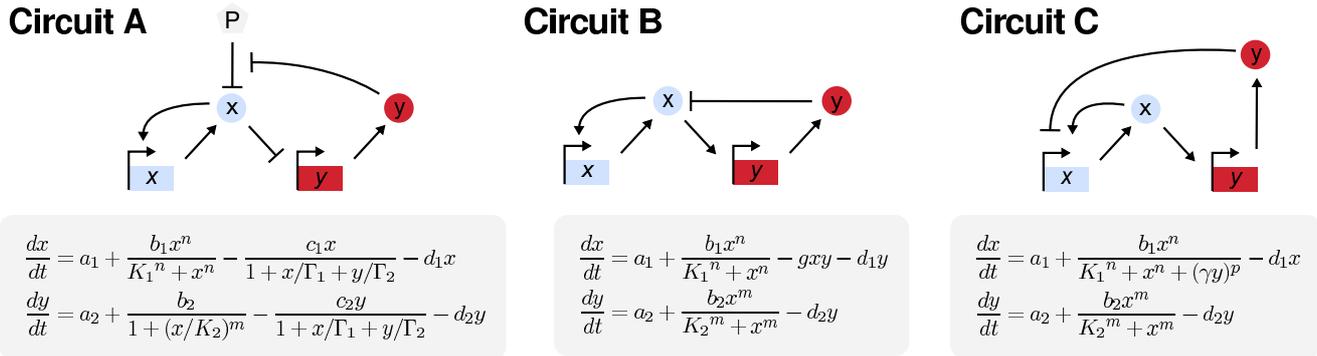}}
\caption{\label{fig:catalogue}
Three excitable circuits (top) and their corresponding continuous mathematical models (bottom). In the circuit
diagrams, uppercase letters and circles represents proteins, and lowercase letters and rectangles denote
genes.}
\end{figure*}

In what follows we will study the three circuits shown in Fig.~\ref{fig:catalogue}.
Circuit A has been identified as the core module underlying transient differentiation
into competence in {\em B. subtilis} cells when placed under nutritional stress \cite{Suel2006}.
A protein \textsf{\textbf{x}}
activates its own transcription and represses the transcription of a second protein
\textsf{\textbf{y}}. The bistable switch on \textsf{\textbf{x}} resulting from the positive
feedback loop is kept off by means of the protease \textsf{\textbf{P}},
which degrades \textsf{\textbf{x}}, but also \textsf{\textbf{y}}. This competitive degradation leads to
excitable behavior, as shown in Sec.~\ref{sec:A}. The continuous description of this
circuit's dynamics in terms of two coupled ordinary equations for \textsf{\textbf{x}} and \textsf{\textbf{y}}
is given below the circuit diagram in Fig.~\ref{fig:catalogue}.

The other two genetic circuits investigated in this work are two-component activator-repressor systems,
in which the activator species forms a 
direct positive feedback loop through its auto-regulation and an indirect negative feedback loop by
means of the activation of its repressor. Specifically, in both circuits B and C
an activator protein \textsf{\textbf{x}} binds to and activates its own expression, and the expression of
a repressor protein \textsf{\textbf{y}}. 
The repressor protein, in turn, inhibits the expression of the activator species
in one of two ways: (i) post-translationally, by inactivating directly the activator protein (degrading it,
for instance), such as in circuit B, or (ii) transcriptionally, by binding
to the promoter of \textsf{\textbf{x}}
and inhibiting its expression, which is the case of circuit C.
The mathematical models of these two
circuits are given below their corresponding diagrams in Fig.~\ref{fig:catalogue}.
These activator-repressor modules are very frequent in natural circuits. In particular, this is a
generic circuit architecture for cell-cycle and circadian oscillators, which need to generate
highly regular periodic activity.
These two circuits have also been implemented
synthetically. The post-translational activator-repressor module (circuit B) has been constructed synthetically in
{\em B. subtilis}, where it has been shown to be functional in the development of
competence \cite{Cagatay2009}, while the transcriptional activator-repressor module (circuit C) was 
built for instance in {\em E. coli} \cite{Atkinson:2003p1217}. Additional synthetic versions
of this latter type of circuit have been built that either include a negative self-feedback of the repressor on itself
\cite{Stricker:2008p373}, or where the repressor action is mediated by quorum sensing
\cite{Danino:2010p447}. 

In the next three Sections we analyze sequentially the potential dynamical regimes that these
circuits exhibit, for different parameter values. As we will see, all three circuit architectures are
able to display the two types of excitable dynamics represented in Fig.~\ref{fig:FIrelation}.
To verify this, we will show the bifurcation diagrams leading to excitability in all cases,
using the unregulated, basal expression of protein \textsf{\textbf{x}} as
bifurcation parameter.
A qualitative comparison between the excitable time traces generated by the discrete stochastic 
simulation of these models in the three circuits will be made.

\section{Circuit A: a competitive degradation circuit}\label{sec:A}

\subsection{Type I excitability close to a saddle-homoclinic bifurcation}\label{sec:circuitA1}

\begin{figure}[htbp]
\centerline{\includegraphics[width=0.45\textwidth]{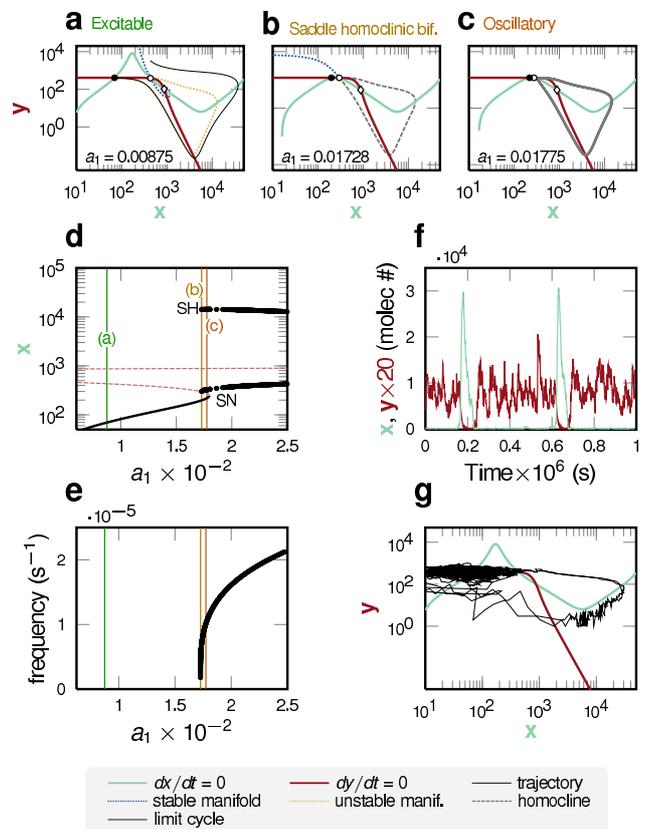}}
\caption{\label{fig:circuitA1}
Type I excitability close to a saddle-homoclinic bifurcation in circuit A. (a-c) Phase
portraits for three values of the parameter $a_1$. The green and red lines represent 
the nullclines of $x$ and $y$. The stable and unstable manifolds of the saddle point
are represented by blue and yellow dotted lines respectively.
(d) Bifurcation diagram of the system for varying $a_1$. Vertical lines
correspond to the values of $a_1$ for which the phase portraits are shown in plots (a-c).
(e) Frequency of the oscillations versus $a_1$. 
(f) Time traces of protein levels for both species in the gene circuit,
as generated from stochastic simulations of its constituent reactions. (g)
Corresponding stochastic trajectory in the phase plane.
The values of the parameters used here are given in Table \ref{tab:params}.
}
\end{figure}

For the set of parameters given in Table~\ref{tab:params} (determined
in a previous work by comparison with experimental observations \cite{Suel2007}),
the competitive degradation circuit labeled
as circuit A in Fig.~\ref{fig:catalogue} exhibits excitable dynamics.
Figure~\ref{fig:circuitA1}(a) shows the phase portrait of the deterministic
model of that circuit (see equations in Fig.~\ref{fig:catalogue}) in
the excitable regime.
\begin{table*}[htbp]
\centering
\small
\begin{tabular}{|c||c|c||c|c||c|c|}
\hline
				&\multicolumn{2}{c||}{\bf Circuit A}			&\multicolumn{2}{c||}{\bf Circuit B}								&\multicolumn{2}{c|}{\bf Circuit C}							\\\hline
				& {\bf Type I}			& {\bf Type II}			& {\bf Type I}					& {\bf Type II}						& {\bf Type I}						& {\bf Type II}			\\\hline\hline
$a_1$			& 0.00875 $\nM/\second$	& 0.016 $\nM/\second$	& 0.00875 $\nM/\second$			& 0.005 $\nM/\second$				& 0.15 $\nM/\second$				& 0.147 $\nM/\second$	\\
$a_2$			& 0 $\nM/\second$		& 0 $\nM/\second$		& 0.075 $\nM/\second$			& 0.025 $\nM/\second$				& 0.05 $\nM/\second$				& 0 $\nM/\second$		\\
$b_1$			& 7.5 $\nM/\second$		& 6 $\nM/\second$		& 7.5 $\nM/\second$				& 15 $\nM/\second$					& 7.5 $\nM/\second$					& 7 $\nM/\second$		\\
$b_2$			& 0.06 $\nM/\second$	& 0.3 $\nM/\second$		& 2.5 $\nM/\second$				& 0.8 $\nM/\second$					& 10 $\nM/\second$					& 0.8 $\nM/\second$		\\
$c_1$			& 0.001 $\nM/\second$	& 0.001 $\nM/\second$	& ---							& ---								& ---								& ---					\\
$c_2$			& 0.001$\nM/\second$	& 0.001	$\nM/\second$	& ---							& ---								& ---								& ---					\\
$d_1$			& 0.0001 $\second^{-1}$	& 0.0001 $\second^{-1}$	& 0.0001 $\second^{-1}$			& $5\cdot 10^{-5}$ $\second^{-1}$	& 0.000386 $\second^{-1}$			& 0.0004$ \second^{-1}$	\\
$d_2$			& 0.0001 $\second^{-1}$	& 0.0001 $\second^{-1}$	& 0.0001 $\second^{-1}$			& $5\cdot 10^{-5}$ $\second^{-1}$	& $3.86\cdot 10^{-5}$ $\second^{-1}$& $4\cdot 10^{-5}$ $\second^{-1}$		\\
$g$				& ---					& ---					& $4\cdot 10^{-8}$ $(\nMs)^{-1}$& $2.5\cdot 10^{-7}$ $(\nMs)^{-1}$	& ---								& ---					\\
$K_1$			& 5000 $\nM$			& 4000 $\nM$			& 5000 $\nM$					& 3000 $\nM$						& 5000 $\nM$						& 5000 $\nM$			\\
$K_2$			& 833 $\nM$				& 500 $\nM$				& 2500 $\nM$					& 750 $\nM$							& 75000 $\nM$						& 2000 $\nM$			\\
$\Gamma_1$		& 25000 $\nM$			& 20000 $\nM$			& ---							& ---								& ---								& ---					\\
$\Gamma_2$		& 20 $\nM$				& 100 $\nM$				& ---							& ---								& ---								& ---					\\
$\gamma$		& ---					& ---					& ---							& ---								& 2									& 0.5						\\
$n$				& 2						& 2						& 2								& 2									& 2									& 2						\\
$m$				& 5						& 2						& 2								& 2									& 2									& 2						\\
$p$				& ---					& ---					& ---							& ---								& 2									& 2						\\\hline
\end{tabular}
\caption{\label{tab:params}
Parameter values used in the deterministic equations of the three gene regulatory circuits. 
For these sets of parameters, the circuits exhibit either type I or type II excitability.
}
\end{table*}
The plot depicts the $x$ and $y$ nullclines (in green and red, respectively, in plots a-c and g), the stable and unstable
manifolds of the saddle, and a sample deterministic trajectory (thin black
line, see figure legend
for a description of the other lines used).
The excitable regime is characterized by a stable fixed point with low levels of \textsf{\textbf{x}}
(black circle) that coexists with two unstable fixed points, a saddle (white circle) and a spiral
(white diamond).
The upper part of the stable manifold of the saddle (blue dotted curve) is the
excitability threshold: any trajectory starting at the right of that manifold is forced to
go through a large excursion in phase space (note the logarithmic scale in the plot's axes)
towards increasing $x$, leading to an excitability pulse.

As the basal expression $a_1$ of \textsf{\textbf{x}} increases, a saddle-homoclinic
bifurcation (SH) takes place, at which the saddle point collides with a stable limit cycle that
surrounds the unstable spiral point (Fig.~\ref{fig:circuitA1}b), giving rise
to a homoclinic orbit (dashed grey line).
For a small parameter range the limit cycle coexists with the stable node
and the saddle (Fig.~\ref{fig:circuitA1}c), until both collide in a saddle-node (SN) bifurcation,
after which the limit cycle (solid grey line) is the only attractor of the system.
This can be seen in the bifurcation diagram shown in Fig.~\ref{fig:circuitA1}(d),
which represents the limit cycle as a thick black line, and the
stable and unstable fixed points as thin solid black and dashed red lines, respectively.

The corresponding behavior of the frequency of the oscillations with respect to
the parameter $a_1$ is represented in Fig.~\ref{fig:circuitA1}(e). The plot
shows that the oscillation frequency at the right of the saddle-homoclinic
bifurcation tends to zero as the bifurcation is approached, which is a clear
hallmark of type I excitable behavior (the frequency values close to zero are
not shown due to numerical constraints). Time traces generated via stochastic
simulations of the set of reactions forming the circuit, using the
standard Gillespie method \cite{Gillespie1977}, are displayed in
Fig.~\ref{fig:circuitA1}(f). The figure shows that the two species taking
part in the dynamics are anticorrelated, with wide pulses of activity of
\textsf{\textbf{x}} leading to dips in the amount of \textsf{\textbf{y}}
protein present in the cell. The corresponding trajectory in
phase space is shown in Fig.~\ref{fig:circuitA1}(g). Previous work \cite{Cagatay2009}
has revealed that this circuit
design leads to highly variable pulse durations, which have the physiological
function of allowing cells to cope with the natural unpredictability
associated with low extracellular DNA levels to which {\em B. subtilis}
cells are subject in the presence of nutritional stress.

\subsection{Type II excitability close to a supercritical Hopf bifurcation}\label{sec:circuitA2}

\begin{figure}[htb]
\centerline{\includegraphics[width=0.45\textwidth]{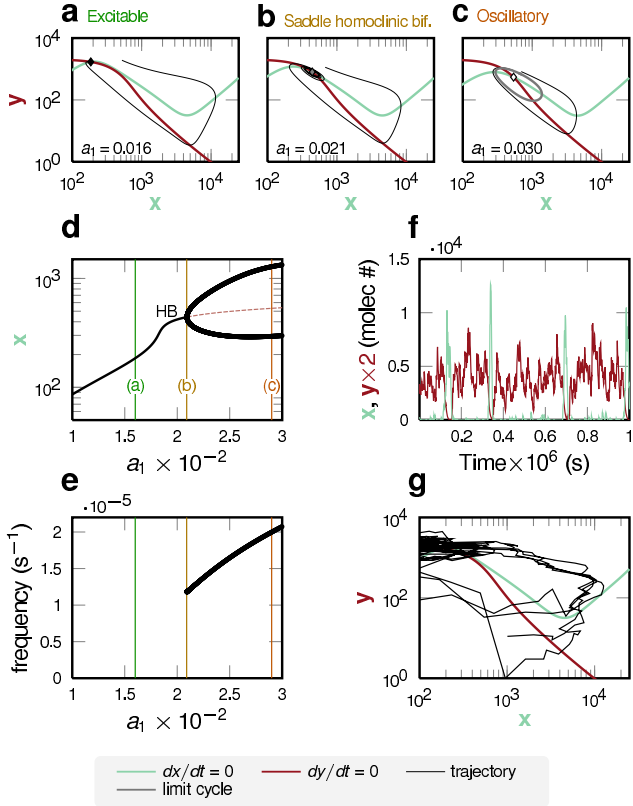}}
\caption{\label{fig:circuitA2}
Type II excitability close to a supercritical Hopf bifurcation in circuit A. The meaning
of the plots is the same as in Fig.~\protect\ref{fig:circuitA1}.}
\end{figure}

A different parameter set (see Table \ref{tab:params}) leads to a completely different
bifurcation scenario that also leads to excitability. The phase portrait of the
system for the excitable regime in this case is shown in
Fig.~\ref{fig:circuitA2}(a). The nullclines now cross only once,
corresponding to a stable spiral point which is the rest state of the excitable
dynamics. Even though there is no saddle here that helps us define an
excitability threshold, a large enough perturbation away from the stable
fixed point is able to elicit a large excursion in phase space, an example
of which is shown as a thin black line in Fig.~\ref{fig:circuitA2}(a).

As the value of $a_1$ increases, the system undergoes a supercritical Hopf bifurcation
(Fig.~\ref{fig:circuitA2}b) that destabilizes the spiral point and generates a stable limit
cycle around it, which is born with zero amplitude, as can be seen in the
bifurcation diagram of Fig.~\ref{fig:circuitA2}(d). Beyond the Hopf bifurcation,
the system exhibits a well-developed stable limit cycle (Fig.~\ref{fig:circuitA2}c).
Interestingly, certain perturbations of this limit cycle evoke large-amplitude excursions
in phase-space, similarly to what happens in the standard excitable regime
shown in Fig.~\ref{fig:circuitA2}(a). An example of such an excursion is shown by a thin solid
line in Fig.~\ref{fig:circuitA2}(c). This regime can be interpreted as
an excitable dynamics with subthreshold oscillations, something that is well-known
to exist in the context of neural systems \cite{Sancristobal:2010uq}.

The frequency response of the circuit to a variation in $a_1$ is shown in
Fig.~\ref{fig:circuitA2}(e). In this case one can see that the frequency remains
finite at the bifurcation, which is characteristic of type II excitability, as mentioned
above. This behavior is a direct consequence of the fact that the underlying
bifurcation bridging excitability and oscillatory dynamics is a supercritical
Hopf bifurcation. Typical time traces of the protein levels of \textsf{\textbf{x}}
and \textsf{\textbf{y}} are shown in Fig.~\ref{fig:circuitA2}(f). Again, as in
the previous Section, there is anticorrelation between the two proteins
during the excitation pulse, although in contrast with the previous case, there
is a large variability in the pulse amplitude. This is due to the fact that the
pulse trajectories in this case depend much more strongly on the initial
conditions of the pulse than in the case of the previous Section
(compare Figs.~\ref{fig:circuitA1}g and \ref{fig:circuitA2}g).

\section{Circuit B: a post-translational activator-repressor circuit}\label{sec:B}

\subsection{Type I excitability close to a SNIC bifurcation}\label{sec:circuitB1}

\begin{figure}[htb]
\centerline{\includegraphics[width=0.45\textwidth]{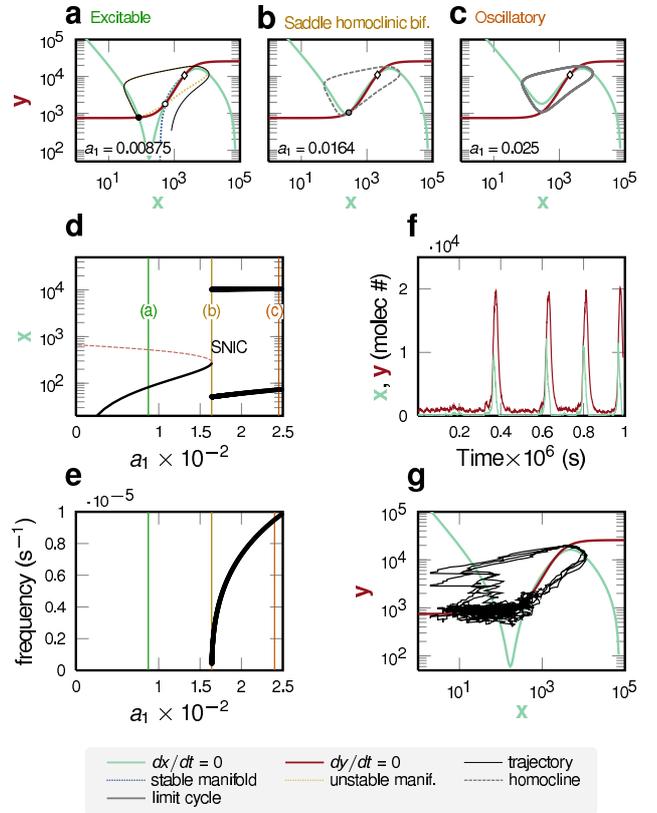}}
\caption{\label{fig:circuitB1}
Type I excitability close to a SNIC bifurcation in circuit B. The meaning
of the plots is the same as in Fig.~\protect\ref{fig:circuitA1}.}
\end{figure}

We now turn to the activator-repressor circuits, beginning with the post-translational
one, in which protein \textsf{\textbf{y}} represses \textsf{\textbf{x}} by
interfering directly, at the protein level, with its activity (for example by
degrading it). For a set of parameters (given in Table \ref{tab:params}) that
successfully reproduce experimental observations \cite{Cagatay2009},
this system also behaves in an excitable manner. The corresponding
phase portrait is shown in Fig.~\ref{fig:circuitB1}(a). Again, as in Sec.~\ref{sec:circuitA1},
the system has three fixed points, only one of which is stable. Perturbations
away from this stable fixed point, that take the trajectory past the stable manifold
of the saddle (which acts again as an excitability threshold), will cause
a large excursion in phase space surrounding the two unstable fixed points,
as represented by a thin black line in the figure.

As the value of the parameter $a_1$ increases, a global bifurcation will
again, as in Sec.~\ref{sec:circuitA1}, generate a limit cycle with a finite
amplitude (Fig.~\ref{fig:circuitB1}b). In contrast with the case of
Sec.~\ref{sec:circuitA1}, however, the bifurcation is now a saddle-node on
an invariant circle (SNIC), since the stable node and the saddle that
exist in the excitable regime collide with each other and annihilate
exactly on top of the newly born limit cycle. The resulting fully developed
limit cycle that exists after the bifurcation is represented in Fig.~\ref{fig:circuitB1}(c),
and the corresponding bifurcation diagram is shown in Fig.~\ref{fig:circuitB1}(d).
The SNIC bifurcation is revealed in this plot by the simultaneous occurrence
of the saddle-node (SN) bifurcation, through which the stable and unstable
fixed-point branches annihilate with each other, and the appearance of the
stable limit cycle, at exactly the same value of $a_1$. 

As summarized in the scheme of Fig.~\ref{fig:FIrelation},
SNIC bifurcations are associated with type I excitable dynamics, for which
the frequency tends to zero at the bifurcation point. This fact is shown
in Fig.~\ref{fig:circuitB1}(e). Time traces and phase-plane trajectories
corresponding to the excitable regime are presented in Figs.~\ref{fig:circuitB1}(f)
and (g). Note that in this case there is a positive correlation between the
two protein species of the circuit, with the repressor slightly tailing behind
(shutting off, in fact) the activator. The positive correlation is a reflection
that the trajectory in phase space is counterclockwise (see Fig.~\ref{fig:circuitB1}a).
Again, as in the case of circuit A (Sec.~\ref{sec:circuitA1}),
there is little dispersion in the pulse amplitudes. In contrast with that other
circuit, however, the dispersion in the pulse widths is known to be here
much smaller \cite{Cagatay2009}.

\subsection{Type II excitability close to a subcritical Hopf bifurcation}\label{sec:circuitB2}

\begin{figure}[htb]
\centerline{\includegraphics[width=0.45\textwidth]{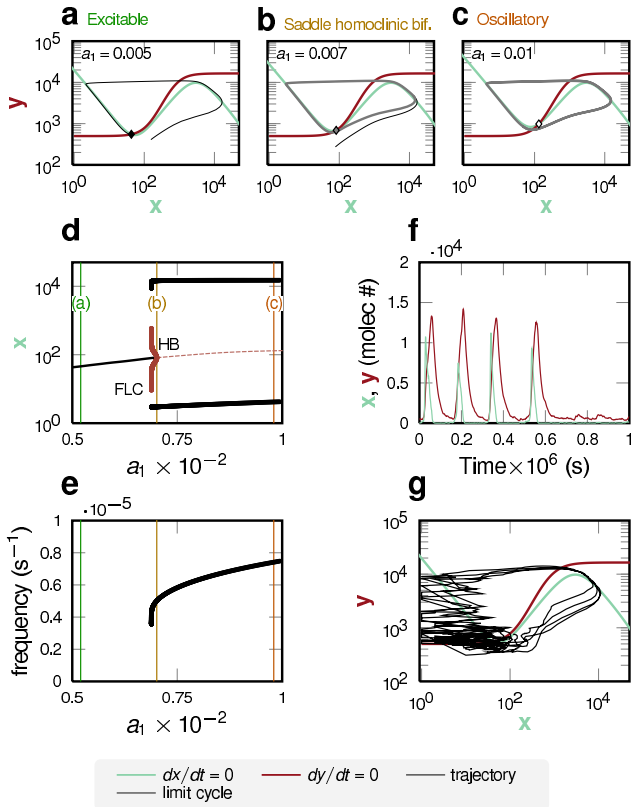}}
\caption{\label{fig:circuitB2}
Type II excitability close to a subcritical Hopf bifurcation in circuit B. The meaning
of the plots is the same as in Fig.~\protect\ref{fig:circuitA1}.}
\end{figure}

The post-translational activator-repressor circuit represented as circuit
B in Fig.~\ref{fig:catalogue} can also exhibit type II dynamics. The
situation is represented in Fig.~\ref{fig:circuitB2}, for which the
corresponding parameters are again given in Table~\ref{tab:params}.

As shown in Fig.~\ref{fig:circuitB2}(a), the excitable regime is characterized by a single fixed point.
The cubic nullcline of the activator and the monotonically increasing nullcline
of the inhibitor make this system analogous to a FitzHugh-Nagumo model.
As in that model,  the dynamics exhibit separation of time scales
between the two species, in such a way that the dynamics of the activator is
much faster than that of the inhibitor.  This is reflected in a quick entry into the
excitable pulse and a slower exit, the latter governed by the final part of the trajectory
in which the system approaches the fixed point again by slowly sliding down the
left branch of the activator nullcline (see Fig.~\ref{fig:circuitB2}a). 
In this case, however, the separation of time scales is not imposed by tuning the
parameters (both species have the same linear degradation terms), but effective and due to 
the post-translational inactivation of \textsf{\textbf{x}} by \textsf{\textbf{y}}.
As the parameter $a_1$
increases, excitability is once more lost, this time via a subcritical Hopf
bifurcation in which the fixed point loses stability via a collision with an
unstable limit cycle (Fig.~\ref{fig:circuitB2}b). This unstable invariant set
was previously generated via a fold limit cycle
(FLC) bifurcation, together with a stable limit cycle which becomes the only
stable attractor (Fig.~\ref{fig:circuitB2}c) after the subcritical Hopf bifurcation occurs.
The situation is summarized in the bifurcation diagram shown in
Fig.~\ref{fig:circuitB2}(d) where the thick red line represents
the unstable limit cycle branch.

The subcritical Hopf bifurcation described above leads to type II dynamics, which
we remind is characterized by the fact that the frequency remains finite at the
bifurcation. This is shown in Fig.~\ref{fig:circuitB2}(e). Finally, Figs.~\ref{fig:circuitB2}(f)
and (g) shows sample trajectories of the system as generated from stochastic
simulations. In this case, as in Sec.~\ref{sec:circuitA2}, the variability of pulse
amplitudes is slightly higher than for type I excitability.

\section{Circuit C: a transcriptional activator-repressor circuit}\label{sec:C}

\subsection{Type I excitability close to a SNIC bifurcation}

\begin{figure}[htb]
\centerline{\includegraphics[width=0.45\textwidth]{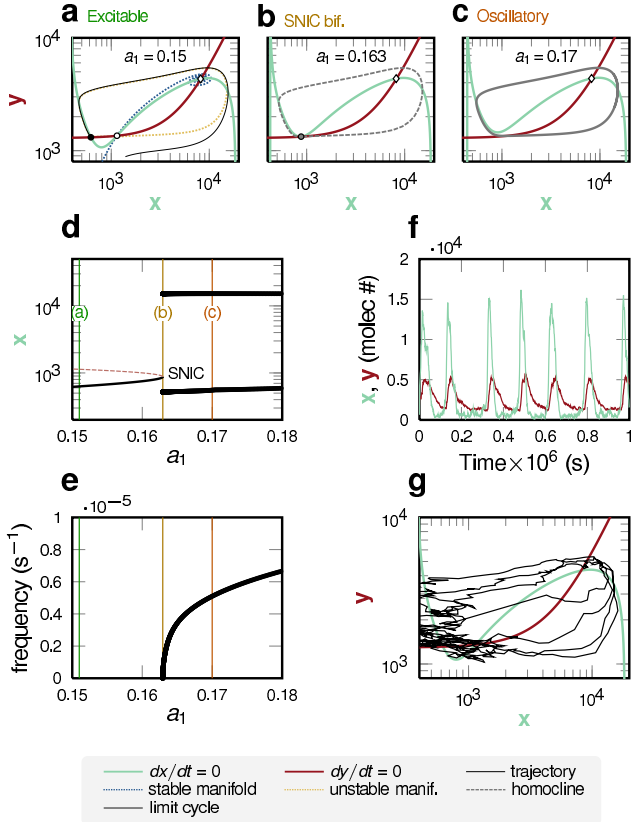}}
\caption{\label{fig:circuitC1}
Type I excitability close to a SNIC bifurcation in circuit C. The meaning
of the plots is the same as in Fig.~\protect\ref{fig:circuitA1}.}
\end{figure}

We now substitute the post-translational repression of \textsf{\textbf{x}} by
\textsf{\textbf{y}} for a transcriptional regulation, in which the repressor
competitively inhibits expression of the activator by binding to its promoter.
In this case the system can also behave as an excitable system, as
shown in Fig.~\ref{fig:circuitC1}(a). As in the previous examples of type I excitability 
discussed above (Secs.~\ref{sec:circuitA1} and \ref{sec:circuitB1}), this system 
has three fixed points, only one of them being stable.
Once more the stable manifold of the saddle defines the excitability threshold,
beyond which a pulse of excitability is triggered. An increase in $a_1$
leads to a SNIC bifurcation in which the stable node and the saddle collide
and annihilate at the same location at which a limit cycle appears
(Fig.~\ref{fig:circuitC1}b). The corresponding bifurcation diagram is shown
in Fig.~\ref{fig:circuitC1}(d).

As expected, the frequency of the limit cycle (shown in Fig.~\ref{fig:circuitC1}c)
decays to zero as the bifurcation is approached, as displayed in
Fig.~\ref{fig:circuitC1}(e). The time traces (Fig.~\ref{fig:circuitC1}f)
and corresponding trajectory in phase space (Fig.~\ref{fig:circuitC1}g)
show that the two species are positively correlated, and that there are slight
variations in the pulse amplitudes (in contrast with the other examples of
type I dynamics discussed above, where the variability in amplitude was basically
absent).

\subsection{Type II excitability close to a subcritical Hopf bifurcation}

\begin{figure}[htbp]
\centerline{\includegraphics[width=0.45\textwidth]{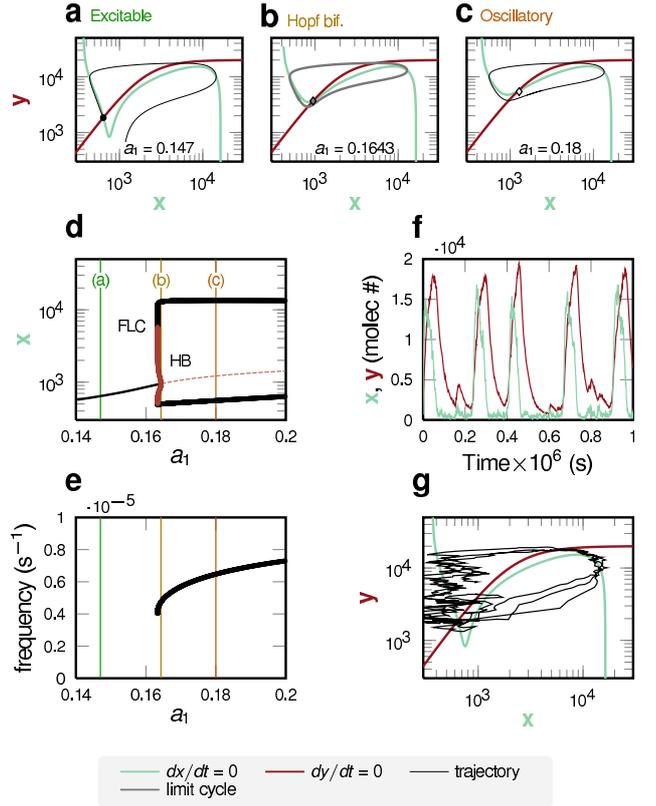}}
\caption{\label{fig:circuitC2}
Type II excitability close to a subcritical Hopf bifurcation in circuit C. The meaning
of the plots is the same as in Fig.~\protect\ref{fig:circuitA1}.}
\end{figure}

Finally, we show that the transcriptional version of the activator-repressor circuit can also
display type II dynamics. Figure~\ref{fig:circuitC2} shows the behavior of the
circuit for the parameters given in Table~\ref{tab:params}. As in the FitzHugh-Nagumo model,
%case of type II excitability in circuit B (Sec.~\ref{sec:circuitB2}), 
we have to assume a clear separation of time scales between the activator and repressor, 
the former being much faster than the latter. The phase portrait corresponding to the excitable 
regime, shown in Fig.~\ref{fig:circuitC2}(a), is again analogous to that exhibited by a
standard FitzHugh-Nagumo model. No well-defined excitability threshold exists,
but perturbations sufficiently far away from the rest state take the system to the
excited branch of the activator nullcline (right branch). The bifurcation is 
very similar to the one exhibited by the post-translational activator-repressor
circuit of Sec.~\ref{sec:circuitB2}. As $a_1$ increases, the rest state loses
stability via a subcritical Hopf bifurcation (Fig.~\ref{fig:circuitC2}b) that
generates a stable limit cycle (Fig.~\ref{fig:circuitC2}c).

The bifurcation plot (Fig.~\ref{fig:circuitC2}d) reveals that the limit cycle is
born via a fold limit cycle bifurcation (FLC) that is closely followed by
the above-mentioned subcritical Hopf bifurcation (HB), after which the
limit cycle is the only stable attractor of the system. Excitable dynamics occurs
before the FLC bifurcation. The period behaves in a manner consistent with 
a type II dynamics, remaining different from zero up to the bifurcation point.
Typical time traces and phase-space trajectories are shown in Figs.~\ref{fig:circuitC2}(f)
and (g), respectively.

\section{Excitability in parameter space}\label{sec:paramspace}

\begin{figure}[htbp]
\centerline{\includegraphics[width=0.45\textwidth]{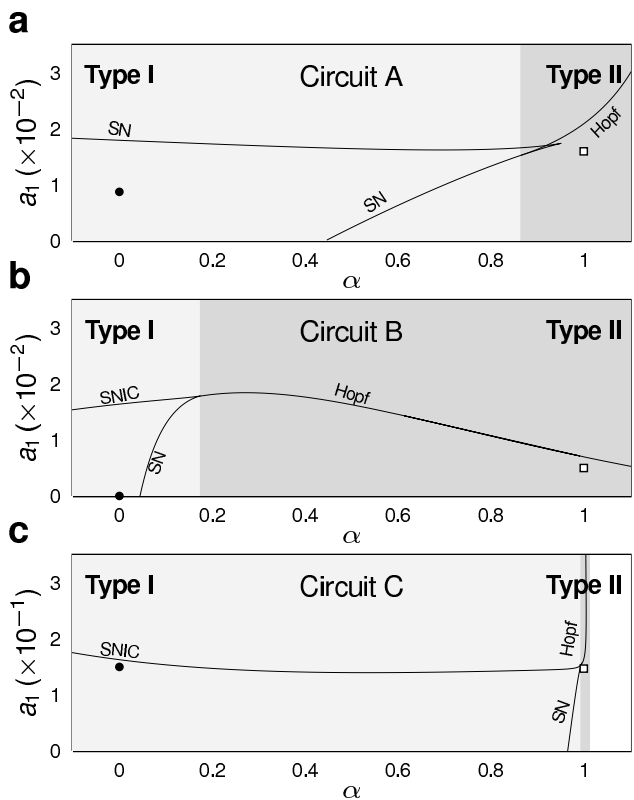}}
\caption{\label{fig:paramspace} Bifurcation lines in the parameter space formed by $\alpha$ and $a_1$ for the three circuits (see main text). 
Clear grey (dark grey) areas correspond to the values of $\alpha$ for which the circuits exhibits type I (type II) excitability.
Black circles denote the sets of parameters used in this text for type I dynamics, and white squares denote parameters for type II behavior.}
\end{figure}

We have shown that all three circuits can exhibit both types of excitability regimes depending 
on specific values of the kinetic parameters.
In this section we show that the parameters used in each case (Table~\ref{tab:params}) are not 
exceptional and that both types of excitability are expected in these circuits. 
For each of the three circuits, we partially explore parameter space between the parameters sets 
used in the previous sections (see Table~\ref{tab:params}).
Specifically, if  $\boldsymbol\pi_{I}$ and  $\boldsymbol\pi_{II}$  are the parameter vectors 
used for type I and type II excitability, we can map a control parameter $\alpha$ to the linear 
combination of these two parameter  vectors, 
$\alpha\mapsto \boldsymbol\pi(\alpha)=(1-\alpha)\cdot\boldsymbol\pi_{I} + \alpha\cdot\boldsymbol\pi_{II}$.
This new parameter allows us to track the transition from type I excitability  at $\alpha=0$ to 
type II excitability at $\alpha=1$, and will give us a sense of how general these regimes are in each of
the circuits studied here.
We have restricted the parameter space scan to values of $\alpha$ for which all the parameters were 
strictly positive.
This restriction defines the range of values that $\alpha$ can take to $(-0.25, 1.667)$ for circuit A, 
$(-0.19, 1.429)$ for circuit B, and $(-27.57, 1)$ for circuit C.
Figure~\ref{fig:paramspace} shows a two-dimensional bifurcation diagram for the parameters $\alpha$ 
and $a_{1}$ for the three circuits.
The transition from type I to type II excitability in circuit A is shown in Fig.~\ref{fig:paramspace}(a).
Type I excitability is maintained as we move from the set of parameters used in Sec.~\ref{sec:circuitA1} 
($\alpha=0$, black circle) towards the set of parameters used in Sec.~\ref{sec:circuitA2} ($\alpha=1$, white square). 
In this Figure, the left-most saddle-node (SN) branch is associated with type I excitable dynamics 
(Fig.~\ref{fig:circuitA1}). 
As $\alpha$ increases, the complementary SN branch appears and both eventually collide in a cusp point. 
Just before this codimension-2 bifurcation occurs, a Bogdanov-Takens bifurcation takes place in the lowest SN 
branch ($\alpha\approx0.865$) and gives rise to the Hopf bifurcation leading to type II excitability.
Thus, in this circuit excitability is preserved for all values of $\alpha$, and both types of excitable dynamics are 
possible in a wide range of parameters.
The same is true for circuit B (Fig.~\ref{fig:paramspace}b) where the SNIC bifurcation responsible of type I 
excitability is annihilated in a cusp point at $\alpha\approx 0.1748$ and the nascent subcritical Hopf bifurcation 
allows again for type II excitability.
Finally, in Circuit C, type I dynamics occupies almost the whole scanned parameter range (see 
Fig.~\ref{fig:paramspace}c), while type II dynamics is relegated to a marginal region. 
In our opinion, this is a limitation of this parameter scanning procedure rather than an imprint of the 
exceptionality of the type II regime in this circuit. The separation of time scales required for the excitability
in this regime plus the large difference in some of the parameter values between the two regimes (e.g. $K_2$), 
makes it difficult to stay in type II excitability while decreasing from $\alpha=1$ to 0.
We have observed that other parameter sets lead to type II dynamics in this circuit (data not shown), 
but this parameter scanning strategy fails in those cases.

\section{Discussion}\label{sec:discussion}

Relating structure and function is an outstanding problem in biology. The design principles
underlying such a mapping can be revealed by means of design space representations \cite{savageau}.
In some cases, multiple circuit architectures can be associated with a single cellular function \cite{sharpe}.
Here we have presented a collection of three gene-circuit designs that are able to exhibit excitable
dynamics. All three architectures are known to exist in nature, and in some cases
excitable dynamics has been observed experimentally. We were interested in determining
whether 
a specific circuit type was necessarily associated with a given type of excitable dynamics,
given that two distinct types are known to exist with different dynamical and statistical
properties of the generated pulses. Our study shows that this is not the case, but that
rather all types of circuits can generate both types of excitable behavior
for a wide range of parameter values (at least in circuits A and B).

This type of design degeneracy can be expected to be very common in natural systems.
Theoretical studies have found this to be the case in noise-buffering circuits
\cite{barkai} and pattern-forming circuits in development, for functions such as
segmentation \cite{ma} and single-stripe formation \cite{sharpe}.
Recent studies \cite{Cagatay2009,Kittisopikul:2010fk} have shown that such a dynamical
degeneracy can be lifted by noise, provided the different circuit architectures involved
are associated with different phase relations between the circuit components: if the circuit topology is
such that the proteins are anti-correlated, one of them will be always occurring in low numbers
and will thus be subject to large fluctuations, so that the overall amount of noise in the system will be high.
On the other hand, if the topology is such that the proteins are correlated, the levels of both proteins will
be high or low simultaneously, and therefore the amount of noise will fluctuate in time
inversely with the protein levels.
In that way, noise
propagates in different ways through the different
circuit designs, and provides the system with specific functional properties
that most likely act as fitness determinants in the evolution of a given molecular
circuitry to perform a desired cellular function \cite{Cagatay2009,Kittisopikul:2010fk}.

On the other hand, the fact that a given circuit can have different dynamical behaviors
would allow evolutionary pressures to shape the specific dynamical response of the cell (for instance in terms of their noise level) without having to modify the circuit's topology.
Hopefully the systematic dynamical 
study presented here could help in establishing how such these selection processes take
place in other excitable cellular functions.

\section*{Acknowledgements}
J.G.O. thanks G\"urol S\"uel, Michael Elowitz, and Alfonso Mart\'inez-Arias for
countless discussions about the role of excitability and bursting
in gene regulatory processes.
This work has been financially supported by the Ministerio de Ciencia e Innovaci\'{o}n
(Spain, project FIS2009-13360 and I3 program)
and the Generalitat de Catalunya (project 2009SGR1168).
P.R. is also supported by the FI programme from the Generalitat de Catalunya.
J.G.O. also acknowledges financial support from the ICREA foundation.

\end{document}